\documentclass[twocolumn]{aastex61}
\pdfoutput=1 
\usepackage{savesym}
\usepackage{xstring}
\usepackage{etoolbox}
\savesymbol{tablenum}
\usepackage{amsmath}
\usepackage[T1]{fontenc}
\usepackage[figure,figure*]{hypcap}
\usepackage{siunitx}
\usepackage{xspace}

\DeclareSIUnit \h {\ensuremath{\mathit{h}}}
\DeclareSIUnit \parsec {pc}\DeclareSIUnit \Msun {M_\odot}
\DeclareSIUnit \particle {part}

\newcommand{\abacus}{\textsc{Abacus}\xspace}
\newcommand{\rockstar}{\textsc{Rockstar}\xspace}
\newcommand{\halotools}{\textsc{halotools}\xspace}
\newcommand{\camb}{\textsc{CAMB}\xspace}
\newcommand{\nbody}{$N$-body\xspace}
\newcommand{\cosmicemu}{\textsc{CosmicEmu}\xspace}
\newcommand{\halofit}{\textsc{HaloFit}\xspace}
\newcommand{\sspline}{\texttt{single\_spline}\xspace}
\newcommand{\bfF}{\mathbf{F}}
\newcommand{\bfr}{\mathbf{r}}
\newcommand{\website}{\url{https://lgarrison.github.io/AbacusCosmos}}
\newcommand{\neff}{\ensuremath{N_\mathrm{eff}}}
\newcommand{\eplanck}[2][]{\texttt{emulator\_\IfEqCase{#2}{%
	{7}{720}%
	{11}{1100}%
}[\PackageError{eplanck}{Undefined option to eplanck: #2}{}]%
box\_planck\ifstrempty{#1}{}{\_#1}}}
\newcommand{\AC}[2][]{\texttt{AbacusCosmos\_\IfEqCase{#2}{%
	{7}{720}%
	{11}{1100}%
}[\PackageError{ac}{Undefined option to AC: #2}{}]%
box\ifstrempty{#1}{}{\_#1}}}

\shorttitle{The \abacus Cosmos}
\shortauthors{Garrison et al.}

\begin{document}

\title{The \abacus Cosmos: A Suite of Cosmological N-body Simulations}

\author{Lehman H.~Garrison}
\affiliation{Harvard-Smithsonian Center for Astrophysics, 60 Garden St., Cambridge, MA 02138}
\author{Daniel J.~Eisenstein}
\affiliation{Harvard-Smithsonian Center for Astrophysics, 60 Garden St., Cambridge, MA 02138}
\author{Douglas Ferrer}
\affiliation{Harvard-Smithsonian Center for Astrophysics, 60 Garden St., Cambridge, MA 02138}
\author{Jeremy L.~Tinker}
\affiliation{Center for Cosmology and Particle Physics, New York University, 4 Washington Place, New York, NY 10003}
\author{Philip A.~Pinto}
\affiliation{Steward Observatory, University of Arizona, 933 N. Cherry Ave., Tucson, AZ 85121}
\author{David H.~Weinberg}
\affiliation{Department of Astronomy, The Ohio State University, 140 West 18th
Avenue, Columbus, OH 43210}
\affiliation{Center for Cosmology and AstroParticle Physics, The Ohio State University, Columbus, OH 43210, USA}

\begin{abstract}
We present a public data release of halo catalogs from a suite of 125 cosmological $N$-body simulations from the \abacus project.  The simulations span 40 $w$CDM cosmologies centered on the Planck 2015 cosmology at two mass resolutions, \SI{4e10}{\per\h\Msun} and \SI{1e10}{\per\h\Msun}, in \SI{1.1}{\per\h\giga\parsec} and \SI{720}{\per\h\mega\parsec} boxes, respectively.  The boxes are phase-matched to suppress sample variance and isolate cosmology dependence.  Additional volume is available via 16 boxes of fixed cosmology and varied phase; a few boxes of single-parameter excursions from Planck 2015 are also provided.  Catalogs spanning $z=1.5$ to $0.1$ are available for friends-of-friends and \rockstar halo finders and include particle subsamples.  All data products are available at \website.
\end{abstract}


\section{Introduction}
High-precision forward modeling of large-scale structure is a necessary complement to upcoming galaxy surveys, including DESI \citep{Levi+2013}, \textit{WFIRST} \citep{Spergel+2015}, \textit{Euclid} \citep{Laureijs+2011}, and LSST \citep{LSST+2009}, that will catalog tens of millions of galaxies in unprecedentedly large volumes.  Mock catalogs must be available that allow design of survey strategies and testing of cosmological parameter estimation and covariance.  Although many fast approximate methods exist for generating such catalogs (e.g.~COLA \citep{Tassev+2013} and QPM \citep{White+2014}; see \cite{Monaco_2016} for a recent review), the highest quality mocks remain those derived from \nbody simulations. These gravity-only simulations directly evolve collisionless Monte Carlo tracers of the matter density field from the early, linear regime to late, non-linear times when galaxies form.  These simulations are fraught with challenges, from bias introduced by discretization at early times \citep[e.g.][]{Joyce_Marcos_2007b,Garrison+2016}, to artificial relaxation at late times \citep[e.g.][]{Diemand+2004,Power+2016}, to lack of hydrodynamical and baryonic physics \citep[][and references therein]{Mead+2015,Schneider_Teyssier_2015,Schaller_2015}, to disagreements among \nbody solvers \citep{Schneider+2016} and among halo finders \citep{Behroozi+2015,Knebe+2013}, to the sheer computational challenge of evolving $\mathcal{O}(\num{e12})$ self-interacting particles, the scale that will be needed within a few years \citep{Schneider+2016,Potter+2016}.  Nevertheless, \nbody remains an invaluable tool in testing models of large-scale structure.

Given the computational expense of \nbody, most public data products lie at one of two extremes: full halo catalogs available at a fixed cosmology, or reduced data products that allow variations in the cosmology.  Some examples of the former are simulations like Bolshoi and MultiDark \citep{Riebe+2013}, Las Damas \citep{McBride+2009}, and Millennium \citep{Lemson+2006}, which provide halo catalogs but focus on a single fiducial cosmology.  Examples of the latter are tools like \cosmicemu \citep{Lawrence+2017} or fitting formulae like those found in \cite{Tinker+2008} or \cite{Comparat+2017} for common data products like the power spectrum or halo mass function.  These tools take cosmological parameters as inputs but do not provide access to the underlying halo catalogs or particles, so the science applications are limited.  We aim to bridge these two extremes by providing a suite of many different cosmologies with access to the underlying halos and subsamples of the particles.  In this sense, our approach is most similar to \url{skiesanduniverses.org} \citep{Klypin+2017} because the focus is access to raw halos and particles.

In the next section, we introduce the code used to run the simulations, and in \S\ref{sec:sim_details} we discuss various parameter choices for the simulations and initial conditions.  In \S\ref{sec:cosm}, we introduce the design of the Latin hypercube grid of 40 cosmologies that form the core of our simulation effort, and in \S\ref{sec:catalogs} we provide an overview of all the available sets of simulations.  In \S\ref{sec:data_products}, we detail the data products that we generate for each simulation, and in \S\ref{sec:validation} we validate their accuracy.  We summarize in \S\ref{sec:summary}.

\section{\abacus: fast and precise $N$-body cosmology}
\subsection{Overview}
The simulations in this work all employ \abacus, an $N$-body code for cosmological simulations first described in \cite{Garrison+2016}.  \abacus is both extremely fast and accurate, capable of computing over 100 billion pairwise force interactions per second on a single computer node, with the option to compute forces to nearly machine precision while maintaining competitive speeds.  It derives its performance from a combination of novel computational techniques and high-performance commodity hardware (in particular GPUs).  \abacus is not yet parallelized for distributed memory systems; each simulation here was run on a single node.  \abacus will be further described in Ferrer et al.~(in prep.) and Metchnik \& Pinto (in prep.); see also \cite{Zhang+2017} for a recent example of a massive, 130 billion particle simulation completed with \abacus.

\subsection{Algorithm}
The computational domain in \abacus is divided into a grid of $\texttt{CPD}^3$ cells, where $\texttt{CPD}$ is the number of cells per dimension.  The force computation is split into near-field and far-field components based on this cell decomposition.  However, unlike most mesh-based $N$-body methods, such as P\textsuperscript{3}M, this decomposition is exact --- the pairwise force between any two particles is always given exactly by either the near-field or far-field force.

Particles interact via the near-field force if they are within \texttt{NearFieldRadius} cells of one another.  For example, \texttt{NearFieldRadius} = 2 means a cell's 124 nearest neighbor cells are included in the near-force calculation.  We compute the force as a direct summation of $1/r^2$ forces (or some appropriately softened form; see below).  This calculation can be accelerated with GPUs, which enables the impressive single-node performance.

Particles interact via the far-field force if they are separated by more than \texttt{NearFieldRadius} cells.  \abacus computes a multipole expansion of the particles in each cell using a variant of the fast multipole method (Metchnik \& Pinto, in prep.; see also \cite{Nitadori_2014} for a similar method).  This multipole expansion is then convolved with a derivatives tensor to yield a set of coefficients for a Taylor-series expansion of the force in a given cell.  The derivatives tensor is a fixed property of the grid for a given \texttt{CPD}, \texttt{NearFieldRadius}, and series expansion \texttt{Order}, and is thus precomputable.  The multipole computation, derivatives convolution, and evaluation of the Taylor series happens for each timestep.  Our performance tuning strategy is to change \texttt{CPD} to balance the cost of the near- and far-field computations since the former can be offloaded to the GPUs while the CPUs are computing the latter
\footnote{Although GPUs are well-suited to convolutions, we find that CPU convolutions are faster for our application when using modern CPUs with large SIMD vectors and a high-performance FFT library like Intel MKL.  This is likely due to the data transfer overhead to the GPU and relatively low compute density (FLOPs per byte) of the operation.}.  The far-field \texttt{Order} parameter largely sets the accuracy; we find $\texttt{Order} = 8$ provides excellent accuracy and performance (see Table \ref{tbl:params}).

The use of the fast multipole method on a discrete mesh enables our exact near-field/far-field force decomposition.  Such an exact decomposition is also possible in tree codes --- one can always choose to open a leaf (near field) or leave it closed and use its multipoles (far field) --- but the key is that all possible vectors from one cell to another are pre-computable if the multipoles are computed on a regular mesh instead of a tree.  This allows one to solve the far-field as a convolution of mesh multipoles instead of explicit multipole interactions.

\subsection{Hardware and performance}
Most of the simulations in this work were run on the El Gato cluster at the University of Arizona.  The GPU nodes are dual-socket machines with two 8-core Intel Ivy Bridge E5-2650v2 processors (2.6 GHz) and 256 GB DDR3 RAM (1800 MHz).  \abacus is not designed to hold all particles in memory but rather stream them from disk, so the I/O demands are fairly high.  In many cases, we use hardware RAID, but for these simulations our strategy was to choose a problem size that would fit on a ramdisk (a filesystem hosted in RAM).  A local scratch disk or network filesystem would typically not have the bandwidth to handle \abacus's I/O demands.  Thus, the choice of $1440^3$ particles was set by the size of the system ramdisk.

Our typical speeds at the outset of this project were about \SI{4}{\mega\particle\per\second} (million particle updates per second) per timestep with about 20\% of this time in the convolution step between each primary timestep; later code improvements increased this to about \SI{12}{\mega\particle\per\second}. This higher speed is still a factor of two lower than typical speeds on our production machines (which were not used in this work), in part due to the extra load on the memory bandwidth from the ramdisk.  The simulations took roughly 1000 timesteps to reach the final redshift from $z_\mathrm{init}=49$.  The large number of timesteps is due to the lack of adaptive timestepping, so the global timestep is set by the shortest dynamical time in the whole box.  This will be addressed with on-the-fly group finding and multi-stepping within those groups in future versions of \abacus.

\subsection{Force softening}\label{sec:softening}
Several force softening options are available in \abacus.  The simplest is Plummer softening, where the $\bfF(\bfr) = \bfr/r^3$ force law is modified as
\begin{equation}\label{eqn:plummer}
\bfF(\bfr) = \frac{\bfr}{(r^2 + \epsilon_p^2)^{3/2}},
\end{equation}
where $\epsilon_p$ is the softening length.  This softening is very fast to compute but is not compact, meaning it never explicitly switches to the exact $r^{-2}$ form at any radius (in contrast with spline softening). This modifies the growth of structure on large scales \citep{Joyce_Marcos_2007b}.  This is the softening we employ for the \eplanck{11} and \eplanck{7} sets of simulations (see \S\ref{sec:catalogs}).

An alternative is spline softening, in which the force law is softened for small radii but explicitly changes to the unsoftened form at large radii.  Traditional spline implementations split the force law into three or more piecewise segments (e.g.~the cubic spline of \cite{Hernquist_Katz_1989}); we split only once for computational efficiency\footnote{Multiple splits can cause code path branching, which incurs a significant performance penalty on CPUs and an even larger one on GPUs.} and call this \sspline.  We derive this form by considering a Taylor expansion in $r$ of Plummer softening (Eq.~\ref{eqn:plummer}) and requiring a smooth transition at the softening scale up to the second derivative\footnote{A Taylor expansion in $r^2$ is also possible, but we discard that solution due to a large plateau of constant angular frequency near $r\sim 0$ that could excite dynamical instabilities.}.  This gives
\begin{equation}
\bfF(\bfr) =
\begin{cases}
\left(10 - 15(r/\epsilon_s) + 6(r/\epsilon_s)^2\right)\bfr/\epsilon_s^3, & r < \epsilon_s; \\
\bfr/r^3, & r >= \epsilon_s.
\end{cases}
\end{equation}
This is the softening we employ for the \AC{11} and \AC{7} simulation sets.

The softening scales $\epsilon_s$ and $\epsilon_p$ imply different minimum dynamical times (an important property, as this sets the requisite temporal resolution to resolve orbits).  We always set the softening length as if it were a Plummer softening and then internally convert to a softening length that gives the same minimum dynamical time for the chosen softening method.  For \sspline, the conversion is $\epsilon_s = 2.16\epsilon_p$.

\section{Simulation Details}\label{sec:sim_details}
We present technical details of the simulation configuration here.  For an overview of the available sets of simulations, see \S\ref{sec:catalogs}.

\subsection{Initial conditions}\label{sec:ics}
The initial conditions were generated by the public \texttt{zeldovich-PLT}\footnote{\url{https://github.com/lgarrison/zeldovich-PLT}} code of \cite{Garrison+2016}.  We do not provide initial conditions files with the catalogs, but we do provide the input parameter file (\texttt{info/abacus.par}) for the IC code and the input power spectrum from \camb \citep{Lewis+1999}.  The initial conditions can thus be generated by re-running the IC code with those inputs.

The simulations use second-order Lagrangian perturbation theory (2LPT) initial conditions, but \texttt{zeldovich-PLT} only outputs first order displacements.  The 2LPT corrections are generated by \abacus on-the-fly using the configuration-space method of \cite{Garrison+2016}.

Two non-standard first-order corrections are implemented by \texttt{zeldovich-PLT}.  The first is that the displacements use the particle lattice eigenmodes rather than the curl-free continuum eigenmodes.  This eliminates transients that arise due to the discretization of the continuum dynamical system (the Vlasov-Boltzmann distribution function) into particles on small scales near $k_\mathrm{Nyquist}$.  The second correction is ``rescaling'', in which initial mode amplitudes are adjusted to counteract the violation of linear theory that inevitably happens on small scales in particle systems.  This violation usually takes the form of growth suppression; thus the initial adjustments are mostly amplitude increases.  We choose $z_\mathrm{target} = 5$ as the redshift at which the rescaled solution will match linear theory; this choice is tested in \cite{Garrison+2016}.

Later, we will refer to simulations that are ``phase-matched'' in the initial conditions.  This refers to initial conditions with the same random number generator seed, called \texttt{ZD\_Seed} in the \texttt{abacus.par} file.  Matching this value (and \texttt{ZD\_NumBlock}) between two simulations guarantees that the amplitudes and phases of the initial modes are identical between the simulations (up to differences in the input power spectrum and cosmology).

\subsection{Input power spectrum}
We use \camb \citep{Lewis+1999} to generate a linear $z=0$ power spectrum for each cosmology in our grid.  We then scale the power spectrum back to $z_\mathrm{init}=49$ by scaling $\sigma_8$ by the ratio of the growth factors $D(z=49)/D(z=0)$.  This $\sigma_8$ is passed to \texttt{zeldovich-PLT}, which handles the re-normalization of the power spectrum.  The computation of the growth factors is done by \abacus's cosmology module, so it is consistent by construction with the simulation's cosmological evolution.  We only use massless neutrinos and include no cosmological neutrino density.  The exact \camb inputs and outputs are available with each simulation.

In the computation of the power spectrum, baryons and CDM are treated as separate species; however, \abacus is a gravity-only $N$-body solver (i.e.~no hydrodynamics or baryonic physics), so we simply combine the baryonic and CDM density when computing the overall mass density of the universe.  It is this combined mass density that sets our particle mass.

\subsection{Code parameters}\label{sec:params}
The important \abacus parameters (including those that might affect code accuracy) are given in Table \ref{tbl:params} and described here.  The simulations were run with a mix of force softening laws.  The ``\texttt{ZD}'' parameter prefix indicates that this is an input to our \texttt{zeldovich-PLT} IC code.

\begin{table}
\centering
\begin{tabular}{l|r}
Parameter & Value \\
\hline
\texttt{LagrangianPTOrder}   & 2 \\
\texttt{Order}               & 8 \\
\texttt{SofteningLength} & 63 or \SI{41}{\kilo\parsec\per\h} \\
\texttt{TimeSliceRedshifts} & 1.5, 1.0, 0.7, 0.5, 0.3, [0.1] \\
\texttt{TimeStepAccel}       & 0.15 \\
\texttt{TimeStepDlna}        & 0.03 \\
\texttt{ZD\_PLT\_target\_z}     & 5 \\
\texttt{ZD\_qPLT}             & 1 \\
\texttt{ZD\_qPLT\_rescale}     & 1 \\
\hline
Max (median) force error & \num{1e-4} (\num{2e-6})
\end{tabular}
\caption{\label{tbl:params} \abacus code parameters, described in \S\ref{sec:params}.  The force error is the maximum fractional error on the unsoftened forces on a set of 66K randomly distributed particles compared to the true $1/r^2$ forces computed with an Ewald summation in 256-bit precision.  The last \texttt{TimeSliceRedshift} is only present for the higher-resolution simulations.}
\end{table}

\begin{description}
\item[\texttt{LagrangianPTOrder}] The order of Lagrangian perturbation theory corrections to compute at runtime (see \S\ref{sec:ics}).
\item[\texttt{Order}] The multipole order used for computing the far-field force.
\item[\texttt{SofteningLength}] The comoving Plummer-equivalent force softening length (see \S\ref{sec:softening}).
\item[\texttt{TimeSliceRedshifts}] The output redshifts.  The last slice ($z=0.1$) is only available for the higher resolution ``\texttt{720box}'' simulations.
\item[\texttt{TimeStepAccel}] Parameter to limit the time step based on particle accelerations.  The time step is set such that $a_\mathrm{max}\Delta t/v_\mathrm{rms}$ is never larger than this parameter. This often sets the choice of time step at late times.  Various choices of this parameter are tested in Ferrer, et al.~(in prep.); $0.15$ is considered a conservative choice.
\item[\texttt{TimeStepDlna}] Maximum $\Delta(\ln a)$ allowed for a time step.  This often sets the choice of time step at early times.  For example, 0.03 means at least 33 steps per $e$-folding of the scale factor.
\item[\texttt{ZD\_PLT\_target\_z}] The target redshift for PLT rescaling of the initial conditions (see \S\ref{sec:ics} for a description of PLT corrections).
\item[\texttt{ZD\_qPLT}] Initialize the displacements and velocities in the eigenmodes of the particle system.
\item[\texttt{ZD\_qPLT\_rescale}] Do PLT rescaling.
\end{description}

\section{Cosmology Grid Design}\label{sec:cosm}
\begin{figure*}[]
\includegraphics[width=.8\paperwidth,trim={0 0 6cm 6cm},clip]{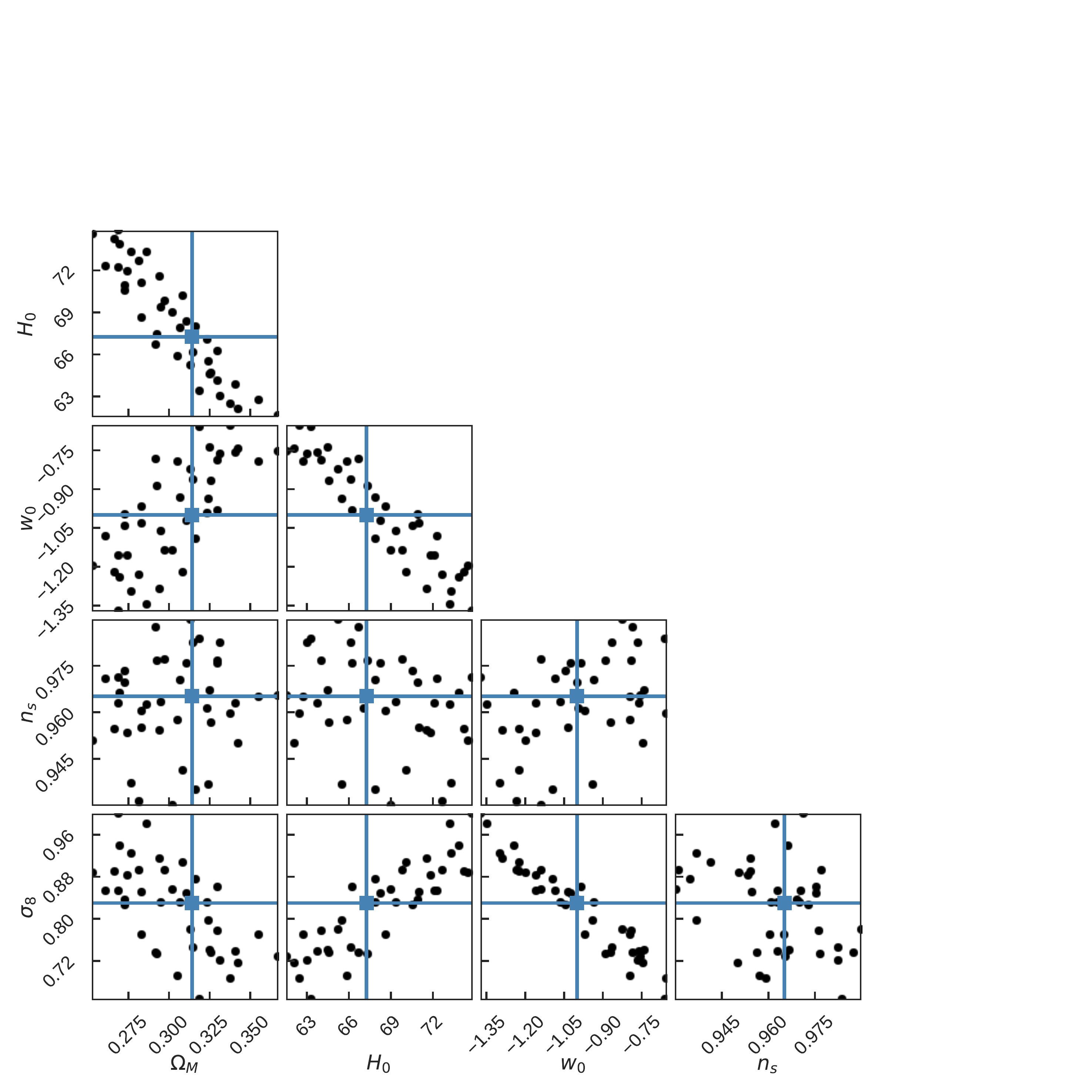}
\caption{A corner plot representation of the cosmology space spanned by the \texttt{AbacusCosmos} simulations, where we have combined $\Omega_\mathrm{CDM}$ and $\Omega_b$ into $\Omega_M$.  The blue square marks the fiducial central cosmology.}
\label{fig:corner}
\end{figure*}

The 40 cosmologies listed in Table \ref{tbl:cosm} are distributed in a 6-dimensional $w$CDM parameter space according to a Latin hypercube algorithm (see Fig.~\ref{fig:corner} for a visual representation).  Specifically, the algorithm samples a $(H_0, \Omega_Mh^2, \Omega_bh^2, \sigma_8, n_s, w_0)$ space centered on the Planck 2013 cosmology \citep{Planck_2013}.  We use $\neff=3.046$ in all cases.  These 40 cosmologies are realized in the two \texttt{AbacusCosmos} simulation sets.

The goal of the cosmological parameter selection is to evenly span the parameter space with only a limited number of cosmologies. The procedure used here follows the Latin hypercube method described in \cite{Heitmann+2009}. In the Latin hypercube method, each of the $N$ dimensions is divided into $M$ bins, with $M$ being the number of desired cosmologies. Each bin in each dimension is only sampled once, thus guaranteeing that the entire range of parameter space is covered in the set of cosmologies. The hypercube is optimized such that, for each cosmology, the distance to the nearest neighboring cosmology is maximized.

The optimized hypercube is then rotated into the parameter space defined by the union of WMAP 9-year \citep{Hinshaw+2013} and Planck 2013 \citep{Planck_2013} CMB results, combined with recent BAO and SN results. The results used to define this space are obtained from \cite{Anderson+2014}. The axes of the Latin hypercube then correspond to the eigenvectors of the CMB-defined parameter space. In the original hypercube design, each axis ranges from 0 to 1. In the CMB parameter space, these ranges correspond to $-4$ to $4$ times the eigenvalue along each eigenvector, so we are sampling from the 4-sigma CMB constraints.

Although the Planck 2013 results were used in the hypercube design instead of the 2015 results, the resulting cosmologies span a larger space than allowed by either data set.  Thus, derivatives measured from these simulations should be equally useful when assuming a fiducial cosmology of either Planck 2013 or 2015.  Indeed, our fiducial cosmology for the \texttt{emulator\_planck} simulations is Planck 2015, and it falls nearly at the center of the cosmology hypercube (the blue square in Fig.~\ref{fig:corner}).

\begin{table*}[p]
\centering
\begin{tabular}{l|rrrrrr}
 &   $H_0$ &   $\Omega_\mathrm{DE}$ &   $\Omega_M$ &    $n_s$ &   $\sigma_8$ &     $w_0$ \\
\hline
 00 & 69   &      0.698 &     0.302 & 0.93  &     0.854 & -1.14  \\
 01 & 63   &      0.669 &     0.331 & 0.982 &     0.719 & -0.765 \\
 02 & 72.3 &      0.739 &     0.261 & 0.97  &     0.851 & -1.08  \\
 03 & 66.2 &      0.671 &     0.329 & 0.975 &     0.858 & -0.982 \\
 04 & 74.2 &      0.733 &     0.267 & 0.954 &     0.889 & -1.22  \\
 05 & 71.5 &      0.706 &     0.294 & 0.954 &     0.913 & -1.29  \\
 06 & 68.6 &      0.717 &     0.283 & 0.96  &     0.768 & -0.969 \\
 07 & 66.7 &      0.709 &     0.291 & 0.987 &     0.734 & -0.788 \\
 08 & 65.8 &      0.695 &     0.305 & 0.957 &     0.692 & -0.796 \\
 09 & 72.7 &      0.719 &     0.281 & 0.931 &     0.89  & -1.24  \\
 10 & 67   &      0.677 &     0.323 & 0.961 &     0.83  & -0.995 \\
 11 & 72.2 &      0.732 &     0.268 & 0.963 &     0.851 & -1.16  \\
 12 & 65.2 &      0.687 &     0.313 & 0.99  &     0.778 & -0.827 \\
 13 & 64.1 &      0.67  &     0.33  & 0.976 &     0.776 & -0.792 \\
 14 & 74.8 &      0.731 &     0.269 & 0.971 &     0.999 & -1.37  \\
 15 & 67.4 &      0.708 &     0.292 & 0.976 &     0.732 & -0.89  \\
 16 & 70.9 &      0.727 &     0.273 & 0.969 &     0.835 & -0.999 \\
 17 & 62.1 &      0.658 &     0.342 & 0.95  &     0.714 & -0.745 \\
 18 & 67.9 &      0.693 &     0.307 & 0.97  &     0.831 & -0.934 \\
 19 & 71   &      0.717 &     0.283 & 0.955 &     0.849 & -1.04  \\
 \end{tabular}
 \begin{tabular}{l|rrrrrr}
  &   $H_0$ &   $\Omega_\mathrm{DE}$ &   $\Omega_M$ &    $n_s$ &   $\sigma_8$ &     $w_0$ \\
\hline
 20 & 64.5 &      0.675 &     0.325 & 0.967 &     0.74  & -0.742 \\
 21 & 68.3 &      0.689 &     0.311 & 0.975 &     0.847 & -1.02  \\
 22 & 73.3 &      0.723 &     0.277 & 0.937 &     0.923 & -1.3   \\
 23 & 62.7 &      0.645 &     0.355 & 0.965 &     0.77  & -0.796 \\
 24 & 73.9 &      0.731 &     0.269 & 0.966 &     0.938 & -1.24  \\
 25 & 61.6 &      0.633 &     0.367 & 0.965 &     0.728 & -0.754 \\
 26 & 69.3 &      0.705 &     0.295 & 0.963 &     0.831 & -1.06  \\
 27 & 62.5 &      0.663 &     0.337 & 0.959 &     0.687 & -0.655 \\
 28 & 65.5 &      0.676 &     0.324 & 0.937 &     0.795 & -0.941 \\
 29 & 67.9 &      0.684 &     0.316 & 0.935 &     0.875 & -1.1   \\
 30 & 64.6 &      0.674 &     0.326 & 0.956 &     0.735 & -0.87  \\
 31 & 69.8 &      0.702 &     0.298 & 0.977 &     0.89  & -1.14  \\
 32 & 63.3 &      0.681 &     0.319 & 0.984 &     0.647 & -0.661 \\
 33 & 66.1 &      0.686 &     0.314 & 0.982 &     0.744 & -0.866 \\
 34 & 70.1 &      0.692 &     0.308 & 0.941 &     0.906 & -1.22  \\
 35 & 73.2 &      0.714 &     0.286 & 0.962 &     0.979 & -1.35  \\
 36 & 63.8 &      0.659 &     0.341 & 0.963 &     0.738 & -0.762 \\
 37 & 70.5 &      0.727 &     0.273 & 0.973 &     0.825 & -1.04  \\
 38 & 74.5 &      0.747 &     0.253 & 0.951 &     0.886 & -1.2   \\
 39 & 71.9 &      0.726 &     0.274 & 0.953 &     0.881 & -1.16  \\
\end{tabular}
\caption{The cosmologies for the \texttt{AbacusCosmos} sets of simulations (both resolutions).  The cosmologies were chosen by a Latin hypercube algorithm centered on the Planck 2013 cosmology; see \S\ref{sec:cosm}. \label{tbl:cosm}}
\end{table*}

\begin{table}
\begin{tabular}{l|rrrrr}
 & $H_0$   & $\Omega_\mathrm{DE}$   & $\Omega_M$   & $\sigma_8$   \\
\hline
 00 \& 00-0 to 00-15    & 67.3 & 0.686      & 0.314     & 0.83      \\
 01    & 67.3    & 0.686          & 0.314         & \textbf{0.78}      \\
 02    & 67.3    & 0.686          & 0.314         & \textbf{0.88}      \\
 03    & \textbf{64.3} & \textbf{0.656}      & \textbf{0.344}     & 0.83      \\
 04    & \textbf{70.3} & \textbf{0.712}      & \textbf{0.288}     & 0.83         \\
\end{tabular}
\caption{The cosmologies for the \eplanck{11} and \eplanck{7} sets of simulations, all with $n_s = 0.965$, $w_0 = -1$, and $\neff = 3.04$.  The first cosmology represents fiducial parameters for which 17 boxes with different phases were run.  The latter four represent ``derivative'' boxes in which one parameter (in bold) is changed at a time.   Note that the cosmologies are actually chosen in the space of physical densities $\Omega_x h^2$, which is why a change in $H_0$ results in change in $\Omega_x$.}
\end{table}

\section{Catalogs}\label{sec:catalogs}
We present four collections of simulations organized into ``sets''.  Sets have names like \AC{11}, while individual simulations have a two-digit number appended to them like \AC[00]{11}.  The two-digit number refers to the cosmology.  Different phases of the same cosmology are indicated by a dash and number after the cosmology, as in \eplanck[00-1]{11}.


The four sets of simulations are as follows:
\begin{description}
\item[\AC{11}] 41 phase-matched%
\footnote{``Phase-matched'' means the initial conditions random number generator was given the same seed. This ensures differences between boxes are due to cosmology and not cosmic variance; see \S\ref{sec:ics}.}
boxes: 40 spanning the 6-dimensional $w$CDM cosmology parameter space (\S\ref{sec:cosm}) and one with the fiducial Planck cosmology.  Each box has size \SI{1100}{\per\h\mega\parsec} and particle mass \SI{4e10}{\per\h\Msun}.  Spline softening of \SI{63}{\per\h\kilo\parsec}.
\item[\AC{7}] Same as the above, but at higher mass resolution (\SI{1e10}{\per\h\Msun}) and smaller box size (\SI{720}{\per\h\mega\parsec}).  Note that these are not zoom-in simulations of the larger boxes, but independent realizations of the power spectrum.  Spline softening of \SI{41}{\per\h\kilo\parsec}.
\item[\eplanck{11}] 21 boxes: 16 with identical Planck cosmologies but different IC phases, and 5 phase-matched boxes of single-parameter variations from the Planck cosmology ($\pm 5\%$ in $\sigma_8$ and $H_0$, plus a central Planck box).  All boxes have size \SI{1100}{\per\h\mega\parsec} and particle mass \SI{4e10}{\per\h\Msun}.  Plummer softening of \SI{63}{\per\h\kilo\parsec}.
\item[\eplanck{7}] Same as the above, but at higher mass resolution (\SI{1e10}{\per\h\Msun}) and smaller box size (\SI{720}{\per\h\mega\parsec}).  As with the \texttt{AbacusCosmos} boxes, these are not zoom-in simulations.  Plummer softening of \SI{41}{\per\h\kilo\parsec}.
\end{description}

The 40 \texttt{AbacusCosmos} simulations are designed to allow estimation of derivatives of cosmological measurables with respect to cosmology.  They can either be used as an ensemble to construct an emulator/interpolator, or each individual box can be differenced with the central \texttt{AbacusCosmos\_planck} box to provide an estimate of the derivative for that particular change in cosmology.

The boxes do not provide much cosmic volume of any individual cosmology (\SI{1.7}{\per\h\cubed\giga\parsec\cubed} if both resolutions are combined), but the 17 phase-varied \texttt{emulator\_planck} boxes provide additional volume and a path toward suppressing cosmic variance and estimating covariance.  The 4 \texttt{emulator\_planck} single-parameter excursion boxes provide a more direct route to measuring derivatives but only for two parameters.

The motivation for two mass resolutions was first to provide a convergence test for large-scale structure properties, at least in the intermediate regime well-sampled by both resolutions (see \S\ref{sec:convergence}).  Second, the larger boxes provide the volume that is needed for BAO-type studies (\citealt{Klypin_Prada_2017} argue that modes longer than \SI{1}{\per\h\giga\parsec} have very little impact on the matter power spectrum), while the smaller boxes provide the halo resolution that is needed by weak-lensing studies \citep[see e.g.][]{Wibking+2017}.

Our softening lengths --- 41 and \SI{63}{\per\h\kilo\parsec} in the two box resolutions --- were chosen to support halos, not subhalos.  Future enhancements to \abacus should make it possible for us to reach dramatically smaller softening lengths.  In the near term, we are continuing to run simulations at these mass scales and resolutions to build volume.  These simulations will be made available on the release website as they finish.

In total, these 125 simulations represent roughly 200K CPU-hours, or 25K GPU-hours, of computational effort.  This is a relatively modest amount for a collection of 370 billion particles and is a testament to \abacus' computational efficiency.

\section{Data products: halos and power spectra}\label{sec:data_products}
We provide three data products at every redshift slice: friends-of-friends (FoF) halos, \rockstar halos, and a high-resolution matter power spectrum.  The redshift slices are $z=\{1.5, 1.0, 0.7, 0.5, 0.3, [0.1]\}$, where the $z=0.1$ slice is only provided for the higher resolution ``\texttt{720box}'' simulations.  Particle subsamples are included with the catalogs.  The data formats of each product are described on the data release website.

\subsection{Friends-of-friends}
The friends-of-friends (FoF) algorithm links particles separated by less than a linking length $b$, equal to $0.186$ in our catalogs (expressed as a fraction of the mean particle spacing).  A halo is defined as a set of linked particles \citep{Davis+1985}.  Our implementation is based on the University of Washington \nbody Shop's halo finder, and we include halos down to 25 particles.  The linking length $0.186$ was chosen to correspond to an overdensity contour of 100 times the background density using the percolation theory results of \cite{More+2011}.

We compute a number of halo properties relative to the FoF centers of mass and velocity, including velocity dispersion, circular velocity profiles, and radial quantiles.  However, the center of mass is susceptible to ``barbell'' pathologies, in which two largely distinct halos are connected by a chance alignment of a thin particle ``bridge''. To guard against this, we also compute a second level of FoF with a smaller linking length (equal to 0.117 in our catalogs).  This linking length was chosen to capture half of the mass of a singular isothermal sphere.  We store the masses of the 4 most massive subhalos and additionally compute global halo properties centered on the most massive subhalo.  These allow for a first-order defense against common FoF pathologies.

We provide a 10\% subsample of particles both inside and outside halos (``halo particles'' and ``field particles'').  Particles are assigned to be subsample particles as a pseudo-random function of the particle ID, so the same 10\% of particles are output at every timestep.  This allows for construction of crude merger trees and other simple box-to-box comparisons.  A uniform sampling of the density field (that is, a 10\% sample of all particles) can be formed via the union of the halo and field particle subsamples.  Particle information includes positions, velocities, and IDs.

\subsection{\rockstar}
\rockstar \citep{Behroozi+2013} is a hierarchical halo finder that uses friends-of-friends in six-dimensional phase space at successively smaller linking lengths to find halos and substructure.  The inner-most substructure defines a halo seed to which particles are assigned based on their phase-space proximity.  \rockstar also can use temporal information (multiple time slices) to improve structure tracking, but we do not use this mode as the time between our outputs is large.  We run \rockstar with mostly default settings, including the default mass definition of $M_{vir}$; however, we report both regular \rockstar masses and strict spherical overdensity masses.  Strict spherical overdensity masses contain all particles, including those considered ``unbound'' and those not associated with the halo.  Subhalos are reported and tagged with their parent halo as decided by \rockstar.  We also output a 10\% subsample of particles in halos.  The exact \rockstar input file is available with each catalog (\texttt{rockstar.cfg}).

\subsection{Power spectra}
We compute the matter power spectra by gridding the particles onto a mesh ($2048^3$ or finer) with triangle-shaped cloud (TSC) mass assignment.  We then Fourier transform the density field, convert the result to a power spectrum, de-convolve the TSC-aliased window function from \cite{Jeong_2010}, and bin in spherical annuli.  The resulting 1D power spectrum is available for every simulation time slice.

\subsection{Plummer vs.~spline data products}
The two \texttt{AbacusCosmos} sets were run with spline softening, while the \texttt{emulator\_planck} boxes were run with Plummer softening (\S\ref{sec:softening}).  The spline was developed partway through the simulation campaign and was thus only applied to the remaining sets of simulations.  However, we also re-ran one of the Plummer boxes with spline softening to calibrate differences in the data products between spline and Plummer.  Those results are presented here.  The simulation is also available on the data release website as \eplanck[spline\_00]{11} if further calibrations are required.

\begin{figure}
\includegraphics[width=\columnwidth]{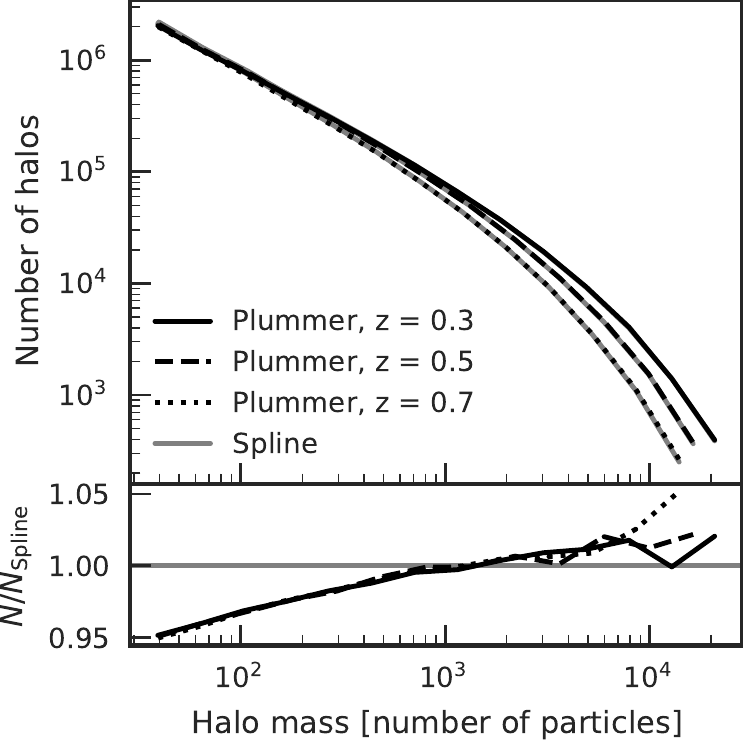}
\caption{Comparison of the FoF halo mass function of two identical simulations that differ only in the force softening technique (Plummer or spline).  For halos larger than 100 particles, the differences are consistently small (< 3\%), although the relative number of Plummer halos steadily increases with increasing mass.  There is almost no evolution of this relation with redshift.}
\label{fig:halo_mass_calibration}
\end{figure}

In Fig.~\ref{fig:halo_mass_calibration}, we show halo mass functions from the friends-of-friends halo finder for Plummer and spline.  The differences are small (< 3\% for halos above 100 particles), but the number of Plummer halos steadily increases with increasing mass (except for the highest mass bin which has very few halos).  The Plummer softening may be inflating halos, causing large halos to come in contact with neighboring structures, increasing their mass.  At the low-mass end, inflating small, tenuously bound halos (below 100 particles) may be causing them to unbind, resulting in fewer halos.  These trends are almost completely insensitive to redshift.  \rockstar and \rockstar SO halos behave very similarly at the low mass end, but Plummer under-predicts the number of $>1000$ particle halos by 1 to 3\%.  This is consistent with the picture that high-mass halos are inflated with Plummer softening, since \rockstar is less susceptible than FoF to over-merging of inflated structures.

\begin{figure}
\includegraphics[width=\columnwidth]{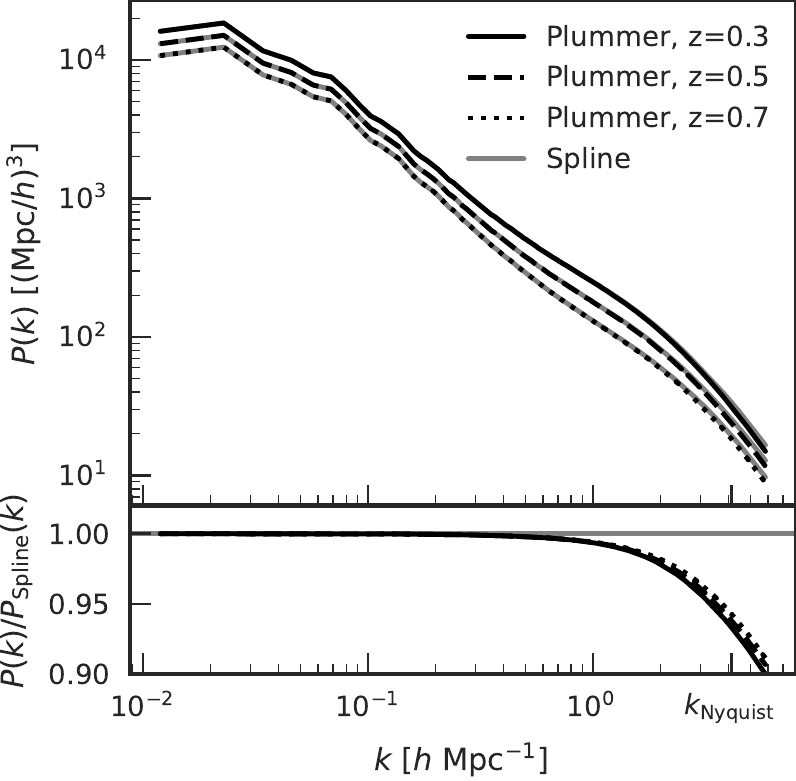}
\caption{Comparison of the matter power spectra of two identical simulations that differ only in the force softening technique (Plummer vs.~spline).  As expected, the Plummer simulation is missing power at high $k$ (6\% at $k_\mathrm{Nyquist}$) due to the long tail of the Plummer force softening law.}
\label{fig:power_calibration}
\end{figure}

In Fig.~\ref{fig:power_calibration}, we show matter power spectra for Plummer and spline softening.  As expected, Plummer results in a loss of power at high $k$.  At the Nyquist wavenumber of the particle lattice, Plummer misses 6\% of power compared to spline.  This loss of power due to non-compact gravitational softening is a known phenomenon; see e.g.~\cite{Joyce_Marcos_2007b,Garrison+2016}.  As with halo mass, there is almost no evolution of this relation with redshift.

\subsection{Example Python Interfaces}
Example Python code is provided on the release website to load and manipulate the halo catalogs, including particle subsamples.  The data formats are also documented on the website, allowing any user to write their own code to parse the catalogs.  However, the provided Python code can at least serve as an implementation example of the data specifications.  A wrapper that loads the halo catalogs into \halotools\footnote{\url{http://halotools.readthedocs.io}} format \citep{Hearin+2017} is also provided.

Another halo occupation distribution (HOD) code that is well-integrated with the Abacus Cosmos simulation suite is \textsc{GRAND-HOD}\footnote{\url{https://github.com/SandyYuan/GRAND-HOD}}.  In particular, \textsc{GRAND-HOD} can use the halo particle subsamples when populating a halo with satellite galaxies.

\section{Validation}\label{sec:validation}
\subsection{\cosmicemu and \halofit}
We validate our power spectrum results against \cosmicemu \citep{Lawrence+2017} which produces a power spectrum as a function of cosmology.  However, the \textsc{AbacusCosmos} cosmologies span a slightly larger parameter space than \cosmicemu, so we are only able to do this test for 28 of our 40 sims.  

\begin{figure}
\includegraphics[width=\columnwidth]{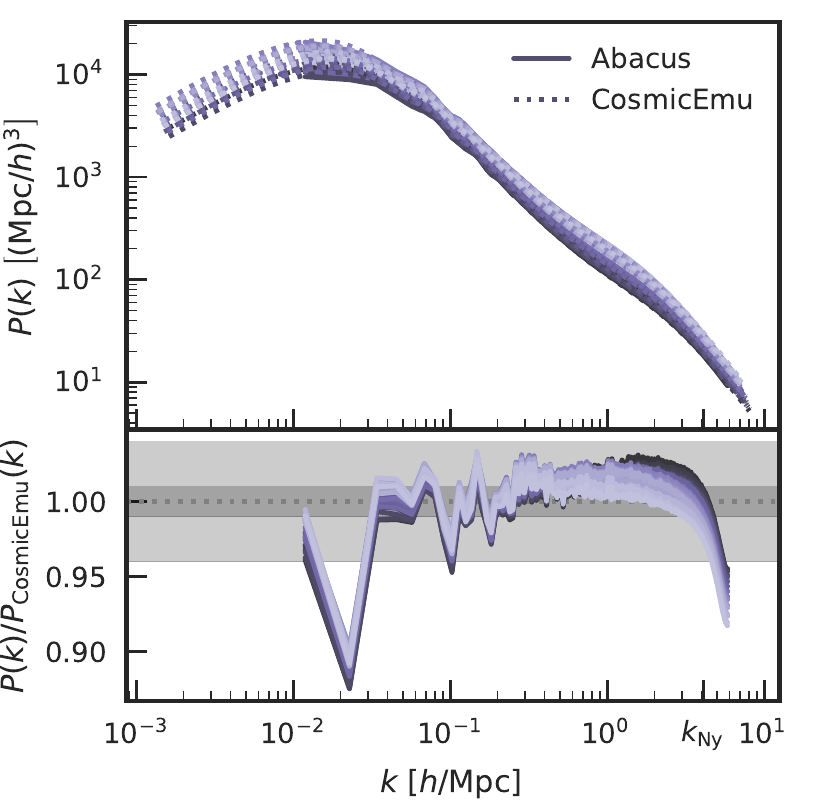}
\caption{Comparison of the $z=0.5$ power spectra from the \AC{11} simulations to the \cosmicemu power spectrum emulator \citep{Lawrence+2017}.  The comparison is shown for the 28 cosmologies that fall within the \cosmicemu domain.  The shaded bar shows the 1\% error region.  Each line represents a simulation; the colors have no meaning beyond distinguishing the lines.  The overall agreement is very good; the low-$k$ differences are due to cosmic variance, while the high-$k$ differences are due to the finite resolution of the simulations.}
\label{fig:AbacusCosmos_validation}
\end{figure}

\begin{figure}
\includegraphics[width=\columnwidth]{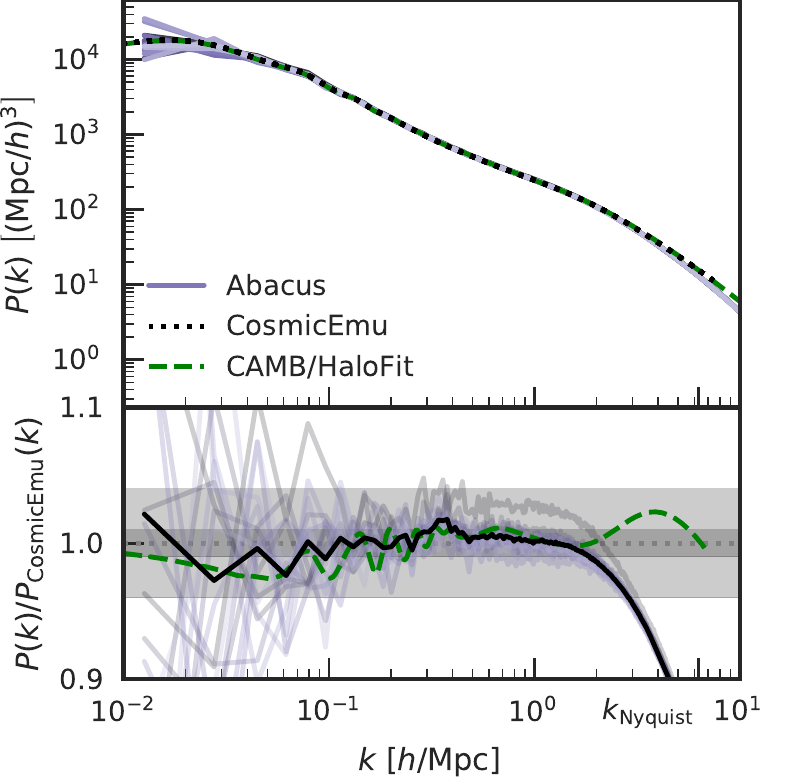}
\caption{Comparison of the $z=0.3$ power spectra from the \eplanck{7} simulations to the \cosmicemu \citep{Lawrence+2017} and \halofit \citep{Takahashi+2012} power spectrum emulators.  The inner shaded bar shows 1\% agreement, while the outer bar shows the 4\% accuracy quoted by \cosmicemu.  Each \abacus line represents a simulation, all of which have the same cosmology but different initial condition phases.  The solid black line is the average of the \abacus lines.  The overall agreement is very good among \abacus, \cosmicemu, and \halofit. The low-$k$ scatter due to cosmic variance is suppressed when averaging over multiple \abacus simulations, while the high-$k$ differences are due to the finite resolution of the simulations and Plummer softening.}
\label{fig:emulator_planck_validation}
\end{figure}

Fig.~\ref{fig:AbacusCosmos_validation} shows that \texttt{AbacusCosmos} and \cosmicemu are in very good agreement (at the level of a few percent) for $k\sim 0.03$ to 4 at z=$0.5$.  At the low-$k$ end, we see large deviations due to cosmic variance, or finite box size\footnote{For statistics that need more cosmological volume, this low-$k$ scatter can be averaged down with the \texttt{emulator\_planck} sims, which have 17 boxes of the same cosmology but different phases.  See Fig.~\ref{fig:emulator_planck_validation}.}.  At the high-$k$ end, we see a downturn in the \abacus power spectra past the Nyquist wavenumber of the particle lattice.  This downturn is expected due to the failure of particle systems reproduce linear theory near, and especially past, $k_\mathrm{Nyquist}$ \citep{Joyce_Marcos_2007b,Garrison+2016}.

Fig.~\ref{fig:emulator_planck_validation} shows the agreement of the \texttt{emulator\_planck} simulations with \cosmicemu and \halofit \citep{Takahashi+2012}, as invoked through \camb's \texttt{do\_nonlinear} feature at $z=0.3$.  In particular, it shows that the low $k$ scatter in the power spectrum seen in Fig.~\ref{fig:AbacusCosmos_validation} can be suppressed by averaging the results of many boxes (recall that these boxes have fixed cosmology but different IC phases).  The agreement is within 4\% (the accuracy quoted by \cosmicemu) for $k\sim 0.01$ to $3$, with a downturn before $k_\mathrm{Nyquist}$ due to the Plummer softening (see also Fig.~\ref{fig:power_calibration}).  The \eplanck[00-0]{7} simulation shows a small systematic deviation around $k=1$ from the mean \abacus result, but we find no evidence for misbehavior of this simulation in our diagnostics.

\subsection{Convergence}\label{sec:convergence}
We compare simulation data products in the intermediate regime well-sampled by both \eplanck{7} and \eplanck{11}.  We examine the power spectrum and halo mass function at three different redshifts.  In all cases, we average over all 17 boxes at each resolution to suppress sample variance.

\begin{figure}
\includegraphics[width=\columnwidth]{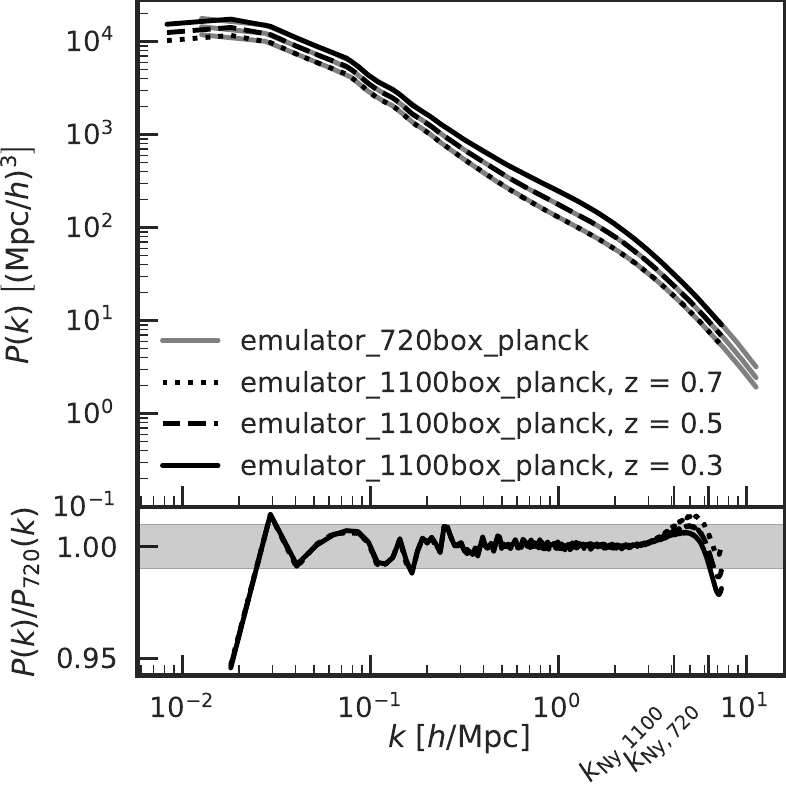}
\caption{A comparison of the matter power spectrum at the two mass resolutions/box sizes.  In the intermediate $k$ regime well-sampled by both resolutions, we expect the results  be converged, since we are averaging over 17 boxes to suppress sample variance.  Indeed, we find excellent agreement of $\sim 1\%$ over a wide range of $k$.  The agreement changes by less than a percentage point from $z=0.7$ to $0.3$.}
\label{fig:power_convergence}
\end{figure}

The matter power spectrum agreement in Fig.~\ref{fig:power_convergence} is excellent from $k=0.03$ to $k=6$ (the Nyquist wavenumber of the higher-resolution box), indicating that our results are stable with respect to box size and mass resolution.  To compare the matter power spectra at incommensurate wavenumbers, we applied a cubic spline in log-log space to the reference spectrum.  However, at the smallest $k$ the binning breaks the smoothness of the power spectrum, so the cubic spline appears to slightly over-predict the disagreement between the resolutions.  Regardless, agreement in this regime is cosmic-variance limited even after averaging over 17 boxes, so this discrepancy is not concerning.

\begin{figure}[!ht]
\includegraphics[width=\columnwidth]{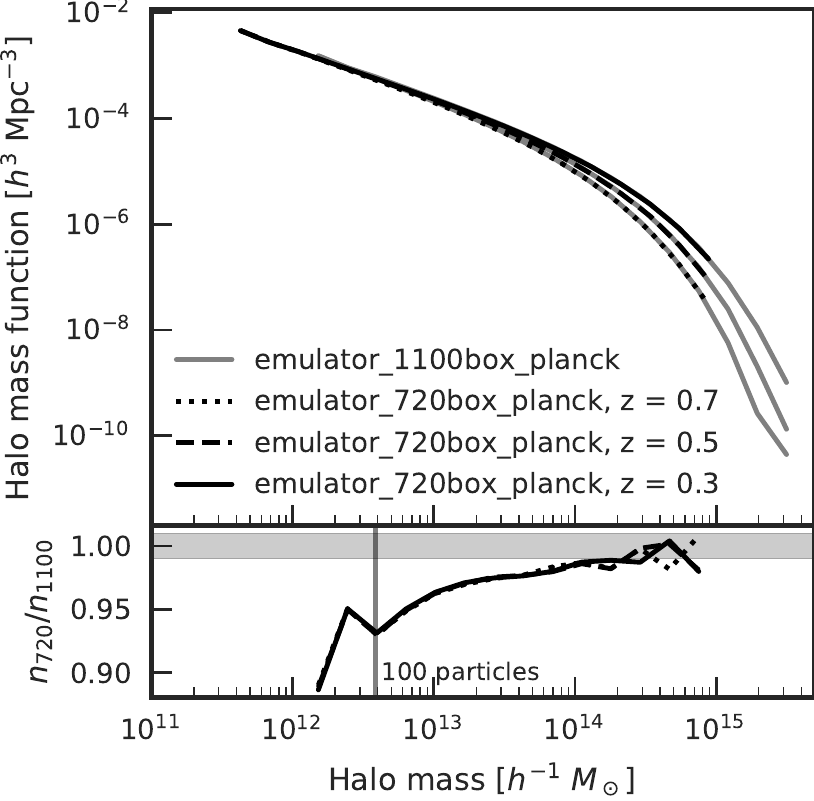}
\caption{A comparison of the FoF halo mass function at the two mass resolutions/box sizes.  The lower mass resolution box over-predicts the number of 100-particle halos by 5\% compared to 400-particle halos at the higher resolution.  This is an effect of friends-of-friends and is not seen with \rockstar; see \S\ref{sec:convergence}.}
\label{fig:hmf_convergence_fof}
\end{figure}

\begin{figure}[!ht]
\includegraphics[width=\columnwidth]{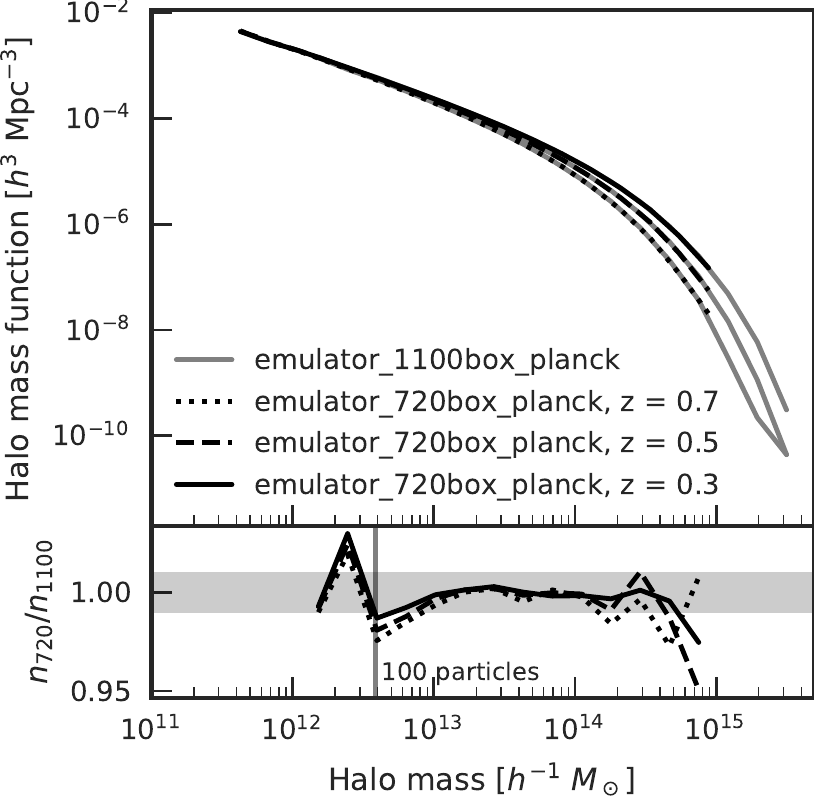}
\caption{Same as Fig.~\ref{fig:hmf_convergence_fof}, but for \rockstar halos.}
\label{fig:hmf_convergence_rockstar}
\end{figure}

The FoF halo mass function is converged to within 6\% in the regime sampled by at least 100 particles at both mass resolutions (Fig.~\ref{fig:hmf_convergence_fof}).  The lower mass resolution systematically overproduces small halos; this is a known effect in friends-of-friends that arises from the mismatch in spatial stochasticity in the particle sampling at the two resolutions \citep{More+2011}.  See figure 10 of \citep{Garrison+2016} for a very similar test, in which downsampling the higher-resolution simulation before running FoF produces excellent agreement with the lower-resolution result.  With \rockstar, the downturn effect almost entirely disappears (Fig.~\ref{fig:hmf_convergence_rockstar}).

\begin{figure}[!ht]
\includegraphics[width=\columnwidth]{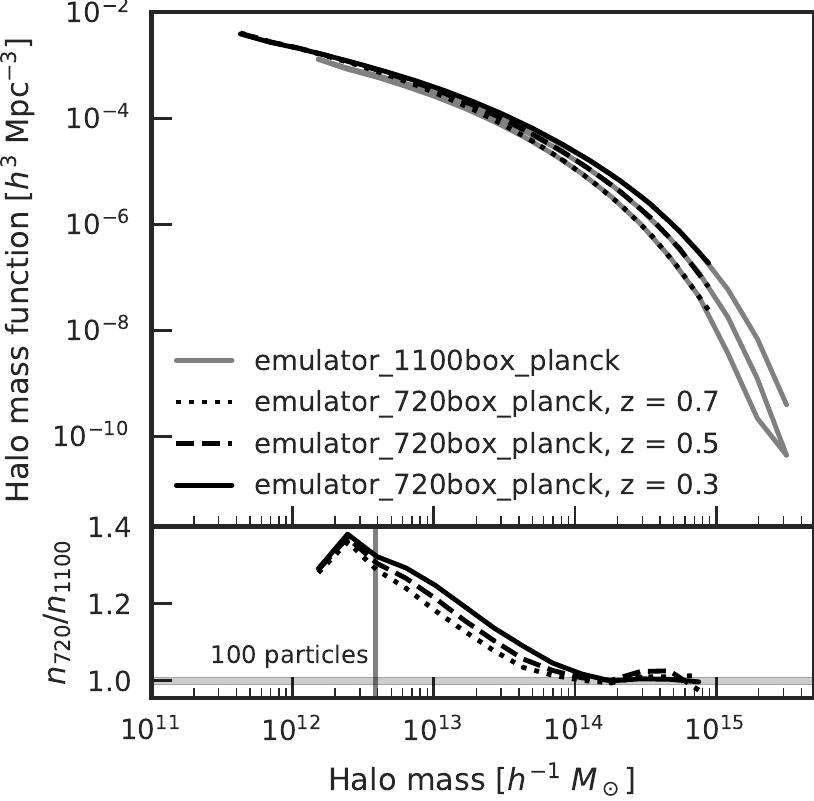}
\caption{Same as Fig.~\ref{fig:hmf_convergence_fof}, but for \rockstar halos with spherical overdensity masses.}
\label{fig:hmf_convergence_rockstar_SO}
\end{figure}

When considering \rockstar SO masses, however, a substantial deviation is seen for all but the largest halos (Fig.~\ref{fig:hmf_convergence_rockstar_SO}).  At 100 particles (400 particles at the higher resolution), 30\% more halos are found at the higher resolution.  We speculate that the higher resolution box finds more physically small halos (due to mass and softening resolution) near large halos; these small halos then have their SO masses inflated by the presence of nearby structure since \rockstar allows overlapping SO spheres and ``double-counting'' of mass.


\section{Summary}\label{sec:summary}
We have presented a suite of cosmological $N$-body simulations produced by the new \abacus code. The modest computational requirements of \abacus (a single GPU node for a few days) enabled us to run one hundred twenty-five $\sim 1$ Gpc boxes, each with 3 billion particles.  These boxes span 40 cosmologies near Planck 2015, allowing for emulation/interpolation in this important parameter region. The accompanying halo catalogs include particle subsamples, allowing for detailed investigations of galaxy bias models and other effects sensitive to dark matter halo structure like weak lensing.  The data products are publicly available and include example code to manipulate the catalogs.

\acknowledgments
We would like to thank the Research Computing Group in the FAS Division of Science at Harvard University for their assistance in hosting the catalogs.  LHG would like to thank Ben Wibking, Andres Salcedo, and Sihan Yuan for their feedback as ``alpha'' testers of these catalogs, Nina Maksimova for helpful discussions on the Abacus algorithm, and the referee for comments that helped improve the quality of the work.  Our FoF implementation is based on the University of Washington N-body Shop friends-of friends code.  This work has been supported by grant AST-1313285 from the National Science Foundation and by grant DE-SC0013718 from the U.S.~Department of Energy.  Some of the computations used in this study were performed on the El Gato supercomputer at the University of Arizona, supported by grant 1228509 from the National Science Foundation.  PAP was supported by NSF AST-1312699.

\bibliographystyle{yahapj}
\bibliography{references}

\end{document}